# Structural Solutions to Dynamic Scheduling for Multimedia Transmission in Unknown Wireless Environments

Fangwen Fu and Mihaela van der Schaar
Department of Electrical Engineering,
University of California, Los Angeles, Los Angeles, CA, 90095
{fwfu, mihaela}@ee.ucla.edu

## ABSTRACT

In this paper, we propose a systematic solution to the problem of scheduling delay-sensitive media data for transmission over time-varying wireless channels. We first formulate the dynamic scheduling problem as a Markov decision process (MDP) that explicitly considers the users' heterogeneous multimedia data characteristics (e.g. delay deadlines, distortion impacts and dependencies etc.) and time-varying channel conditions, which are not simultaneously considered in state-of-the-art packet scheduling algorithms. This formulation allows us to perform foresighted decisions to schedule multiple data units for transmission at each time in order to optimize the long-term utilities of the multimedia applications. The heterogeneity of the media data enables us to express the transmission priorities between the different data units as a priority graph, which is a directed acyclic graph (DAG). This priority graph provides us with an elegant structure to decompose the multi-data unit foresighted decision at each time into multiple single-data unit foresighted decisions which can be performed sequentially, from the high priority data units to the low priority data units, thereby significantly reducing the computation complexity. When the statistical knowledge of the multimedia data characteristics and channel conditions is unknown a priori, we develop a low-complexity online learning algorithm to update the value functions which capture the impact of the current decision on the future utility. The simulation results show that the proposed solution significantly outperforms existing state-of-the-art scheduling solutions.

## I. INTRODUCTION

Existing wireless networks provide dynamically varying resources with only limited support for the Quality of Service (QoS) required by *delay-sensitive, bandwidth-intense and loss-tolerant* multimedia applications. One of the key challenges for delivering the multimedia data over wireless networks is the *dynamic* characteristics of both the wireless channels and the multimedia source data [1]. To overcome this challenge, packet scheduling optimization has been extensively investigated in recent years in order to maximize the quality of the multimedia application given the underlying resource constraints. A brief summary of the existing research on packet scheduling is provided in Table 1.

Table 1. Summary of various existing packet scheduling methods

| Methods | Performance metric | Heterogeneous traffic characteristics | Time-varying channel conditions | Transmission acknowledgement[1] |
|---|---|---|---|---|
| [2][5] | Average delay | No | Yes | Immediate |
| [3][4] | Consumed energy | Hard delay deadlines | No | Immediate |
| [8][9][25] | Application distortion | Packet distortion impacts | No | Immediate |
| [6][11] | Application distortion | Packet distortion impacts, hard delay deadlines, packet dependencies | Yes | Immediate |
| [7][10][12] | Application distortion | Packet distortion impacts, hard delay deadlines, packet dependencies | No | Delayed |
| [13] [26] | Application distortion | Packet distortion impacts, hard delay deadlines | Yes | Immediate |
| Proposed | Application distortion | Packet distortion impacts, hard delay deadlines, packet dependencies | Yes | Immediate |

The multimedia data is often encoded and encapsulated into multiple data units (DUs), which can be video frames, packets etc. Different DUs often have different distortion impacts, delay deadlines and dependencies. Existing packet scheduling solutions often ignore these heterogeneous characteristics of multimedia applications. For example, in [2][5], a packet scheduling method is proposed for minimizing the incurred average delay under energy (or average power) constraints for homogeneous applications in which the packets are not differentiated. In [4], the optimal packet scheduling algorithm is developed for the transmission of a group of equally important packets, which minimizes the consumed energy while satisfying their common delay deadline. The packet scheduling algorithm in [4] is extended in [3] to the case in which each packet has its own individual delay constraint. However, the above papers disregard key properties of multimedia applications: the interdependencies among packets and their different distortion impacts.

In [8][9][25], the packets are scheduled for transmission over a constant channel (with constant packet error rate) in order to minimize the application distortion while satisfying the imposed delay constraints. However, these solutions do not take into account the complicated dependencies between media packets.

---

[1] Immediate acknowledge means that the acknowledgement of one packet transmission is received before next packet starts to transmit.

In [26], a channel, deadline, and distortion-aware packet scheduling algorithm is developed, where only one frame is considered for transmission at each time and an i.i.d. channel is assumed. In this paper, the dependencies between video frames are not considered. In [13], the packet scheduling is optimized based on packets' hard delay deadlines, distortion impacts and the underlying time-varying wireless channel. However, [13] did not explicitly take into account the dependencies between packets. To take into account the dependencies between packets, in [7][10][12], the packet scheduling is optimized using a rate-distortion framework (named RaDiO), which expresses the dependencies between packets as a DAG. However, RaDiO disregards the time-varying characteristics of the considered transmission network, thereby leading to a suboptimal performance over wireless networks. In [6][11], the scheduling of video packets over a time-varying wireless channel is formulated as a cross-layer optimization problem. However, these cross-layer optimization solutions only maximize the quality of the currently transmitted video packets based on the observed channel conditions, without considering future transmission opportunities and the impact of current decisions on the long-term video quality. This type of optimization will be referred to in our paper as myopic optimization.

In summary, a systematic packet scheduling optimization framework for media communication that explicitly considers both the heterogeneous characteristics of the multimedia traffic and the time-varying wireless conditions is still missing. To overcome this challenge, in this paper we develop a systematic energy-aware packet scheduling framework for single-user multimedia transmission over a time-varying wireless channel.

To capture the heterogeneous characteristics of DUs, we first introduce the concept of a "context" at each time slot to represent the heterogeneity of the DUs available for transmission at each time slot. Through the context concept, we are able to capture the dynamic features of the multimedia packets across time. We then formulate the dynamic packet scheduling optimization as a Markov decision process (MDP) problem [14] by further considering the underlying channel dynamics. Within the MDP formulation, the packet scheduling is performed in a foresighted fashion in order to maximize the long-term reconstructed multimedia quality.

In the conventional MDP formulation, the foresighted decision for the packet scheduling is often coupled with the expectation over the experienced dynamics, which makes the packet scheduling problem hard to solve in unknown environments (i.e. where the statistical knowledge of the multimedia data arrivals and channel state transitions is unknown). To resolve this obstacle, we introduce a post-decision state (which is a "middle" state, in which the transmitter finds itself after packet transmission but before the new packet arrivals and new channel realization) and a corresponding post-decision state-value function which represents the optimal long-term utility starting from the post-decision state. Through the post-decision state value function, we can *separate* the foresighted decision on the packet scheduling from the expectation over the underlying dynamics. In other words, the foresighted decision can be computed without knowing the experienced dynamics, given the post-decision state value function. The post-decision state value functions can then be updated online accordingly. Hence, the post-decision state concept allows us to separate the packet scheduling at each time slot into two phases (i.e. *two-phase packet scheduling*): one is the foresighted decision on the optimal scheduling given the post-decision state value function, and the other one is the online update on the post-decision state value function.

In order to reduce the complexity involved in computing the packet scheduling policy, we define the transmission priorities of the DUs in each context based on the distortion impacts, delay deadlines and dependencies, and express them as a DAG, which we refer to simply as the priority graph. Different from the DAG expression on the source coding dependencies in [7], the proposed DAG construction represents the transmission priorities which include the packet dependencies. Through the priority graph, we are able to *separate* the multi-DU foresighted decision at each time slot into multiple single-DU foresighted decisions and the two-phase packet scheduling is applied to each individual DU, which significantly reduce the complexity in computing the optimal foresighted decisions.

In the unknown environment, we further develop an online learning algorithm to estimate the post-decision state value functions. Based on the separation developed for multi-DU foresighted optimization, we are able to estimate the post-decision state value functions for each DU using a low-complexity online learning method.

The paper is organized as follows. Section II characterizes the attributes of the multimedia traffic. Section III formulates the packet scheduling problem for multiple independently decodable DUs as an MDP and develops structural solutions to determine the optimal packet scheduling policies. Section IV further extends the structural results to the packet scheduling for interdependent DUs. Section V presents the simulation results, followed by the conclusions in Section VI.

## II. MULTIMEDIA TRAFFIC CHARACTERISTICS

In this section, we discuss how the heterogeneous attributes of multimedia traffic[2] can be represented. In past work, multimedia traffic (e.g. video traffic) is often modelled as a leaky bucket with constraints (e.g. peak rate constraint, average delay constraint etc.) [19]. However, this model only characterizes the rate change in multimedia traffic and does not explicitly consider the heterogeneous characteristics of multimedia data. In this section, we aim to develop a general model to represent the encoded multimedia data with heterogeneous characteristics (e.g. various delay deadlines, distortion impacts, dependencies, etc.). Using this multimedia traffic model, we will be able to dynamically optimize packet scheduling for multimedia transmission over time-varying wireless networks, which is presented in Sections III and IV.

### A. Attributes of data units

In this section, we discuss how the heterogeneous attributes of the multimedia data can be modelled. The multimedia data is often encoded periodically using a Group of Pictures (GOP) structure, which lasts a period of $T$ time slots. The multimedia data within one GOP are encoded interdependently using, e.g. motion estimation, while the data belonging to different GOPs are encoded independently. Note that the prediction-based coding schemes often lead to sophisticated dependencies. After being encoded, each GOP contains $N$ data units (DUs), each representing one type of DU (e.g. I, P, B frames in encoded video bitstream) and being indexed by $j \in \{1,\cdots,N\}$. The set of DUs within GOP $g \in \mathbb{N}$ is denoted by $\{f_1^g,\cdots,f_N^g\}$. The attributes of DU $f_j^g$ are listed below.

*Size:* The size of DU $f_j^g$ is denoted by $l_{f_j^g}$ (measured in packets[3]), where $l_{f_j^g} \in [1, l_j^{\max}]$, and $l_j^{\max}$ is the maximum allowable size for the $j$-th DU at each GOP. The size of DU $f_j^g$ is determined when DU $f_j^g$ is encoded. To simplify the exposition, $l_j^g$ is generated from an i.i.d. random variable[4] with the probability mass function $PMF_{f_j^g}(l)$. Note that $PMF_{f_j^g}(l)$ is the same for the $j$-th DU across different GOPs, but it differs for different types of DUs.

---

[2] Multimedia traffic can be generated in real time or be pre-encoded.
[3] For simplicity, we assume in this paper that each packet has the same length, but this does not affect our proposed solution. It just simplifies our exposition given the space limitations.
[4] The DU size can also be modeled as a random variable depending on the previous DUs.

*Distortion impact*: Each DU $f_j^g$ has a distortion impact $q_{f_j^g}$ per packet, which is assumed to be the same for all the GOPs, i.e. $q_{f_j^g} = q_{f_j^{g'}}, \forall g, g'$. The distortion impact $q_{f_j^g}$ represents the amount by which the multimedia distortion is reduced if one packet from DU $f_j^g$ is received at the decoder side.

*Delay deadline*: The delay deadline of DU $f_j^g$ represents the time by which the DU should be decoded in order to be displayed. We denote by $d_{f_j^g}$ the delay deadline of DU $f_j^g$. Since the GOP structure is fixed, the difference between the delay deadlines of the two DUs within one GOP is constant, i.e. $d_{f_j^g} - d_{f_{j'}^g} = \Delta d_{jj'} > 0$ where $j > j'$, and the delay deadlines of the same types of DUs from different GOPs satisfy $d_{f_j^g} - d_{f_j^{g-1}} = T$. In other words, the $j$-th DU periodically appears at each GOP with the period of $T$ time slots, which is the length of one GOP.

*Dependency*: When one DU $f_j^g$ is encoded based on the prediction from the other DU $f_{j'}^g$, we say that DU $f_j^g$ depends on DU $f_{j'}^g$. Note that the dependencies between DUs only occur within one GOP and DUs from different GOPs can be decoded independently (i.e. no dependency between them.). The dependencies between the DUs within one GOP are expressed as a directed acyclic graph (DAG) [7]. The DAG remains the same for a fixed GOP structure. In this paper, we assume that, if DU $f_j^g$ depends on DU $f_{j'}^g$ (i.e. there exists a path directed from DU $f_j^g$ to DU $f_{j'}^g$ in the DAG and denoted by $f_{j'}^g \prec f_j^g$), then $d_{f_j^g} \geq d_{f_{j'}^g}$ and $q_{f_j^g} \leq q_{f_{j'}^g}$. In other words, DU $f_{j'}^g$ should be decoded prior to DU $f_j^g$ and DU $f_{j'}^g$ has higher distortion impact.

## B. Traffic state representation in each time slot

We consider a time-slotted system in which the $n$-th time slot is defined as the time interval $[n\Delta t, (n+1)\Delta t)$, where $\Delta t$ is the length of one time slot. In this subsection, we discuss how to represent the multimedia traffic which is ready for transmission at each time slot. At time slot $t$, as in [7], we assume that the wireless user will only consider for transmission the DUs with delay deadlines in the range of $[t, t+W)$, where $W$ is referred to as the scheduling time window (STW) and assumed to be determined a priori[5]. In this paper, we further assume that the STW is chosen to satisfy the following condition: if DU $f_j^g$ directly depends on DU $f_{j'}^g$ (i.e. there is a direct arc from $f_j^g$ to $f_{j'}^g$ in the DAG), then $d_{f_j^g} - d_{f_{j'}^g} < W$. This assumption ensures that DU $f_j^g$ and $f_{j'}^g$ can be considered for transmission at the same time slot.

At time slot $t$, we introduce a context to represent the set of DUs that are considered for transmission, i.e. whose delay deadlines are within the range of $[t, t+W)$ [6]. We denote the context by $C_t = \{f_j^g \mid d_{f_j^g} \in [t, t+W)\}$. Since the GOP structure is fixed, it is easy to show that $C_t$ is periodic with the period of $T$, which means that, for any $f_j^g \in C_t$, there exists $f_j^{g+1} \in C_{t+T}$ and vice versa. Hence, $C_t$ and $C_{t+T}$ have the same types of DUs and the same DAG between these DUs. For example, as shown in Figure 1, $C_t = \{f_1^g, f_2^g, f_3^g\}$ and $C_{t+3} = \{f_1^{g+1}, f_2^{g+1}, f_3^{g+1}\}$, where $T = 3$. Since the context represents the set of DUs to be transmitted, it implicitly represents the dependencies between the DUs. Due to the periodicity, there are only $T$ different contexts. The transition from context $C_t$ to $C_{t+1}$ is deterministic. It is worth to know that, the context indicates the distortion impacts of the DUs and the dependencies between DUs and the context transition indicates the delay deadlines of the DUs.

Given the current context $C_t$, we let $x_{f,t}$ denote the number of packets in the buffer associated with DU $f \in C_t$. Note that $x_{f,t} \leq l_f$, where $l_f$ represents the amount of the originally available packets for DU $f$. We denote the buffer state of the DUs in $C_t$ by $\boldsymbol{x}_t = \{x_{f,t} \mid f \in C_t\}$. The traffic state $T_t$ at time slot $t$ is then defined as $T_t = (C_t, \boldsymbol{x}_t)$, where the context represents the types of DUs, the dependencies between them, and the buffer state $\boldsymbol{x}_t$ represents the amount of packets remaining for transmission. Hence, the traffic state $T_t$ is able to capture heterogeneous multimedia traffic and is a super-set of existing well-known single-buffer models (i.e. which ignore packet dependencies and delay deadlines) or multi-buffer models (i.e. which ignore packet dependencies *or* delay deadlines).

---

[5] The STW can be determined based on the channel conditions experienced by the user in each time slot. For example, the STW can be set small when the channel conditions are poor (i.e. in the low SNR regime), and large whenever the channel condition are good (i.e. in the large SNR regime).

[6] We assume that $d_j - t_j \geq W$ which means that the DUs that are considered for transmission at the time slot $t$ must arrive no later than time slot.

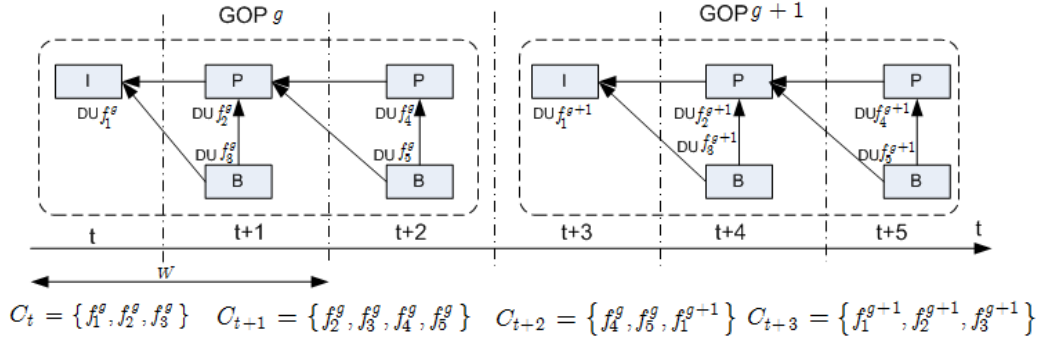

Figure 1. . DAG-based dependencies and traffic states at each time slot using IBPBP GOP structure

## III. PACKET SCHEDULING FOR INDEPENDENTLY DECODABLE DUs

We first consider how the packet scheduling optimization should be performed for the DUs that can be independently decoded (e.g. motion JPEG), and consider the interdependent DUs in Section IV.

At each time slot $t$, the wireless user experiences a channel condition $h_t \in \mathcal{H}$, where $\mathcal{H}$ is the set of finite possible channel conditions and $h_t$ is referred to as the channel state. In this paper, we assume that the channel condition $h_t$ can be modelled as a finite-state Markov chain (FSMC) [18] with transition probability $p_h(h' \mid h) \in [0,1]$. We further define the state which the wireless user experiences at each time slot $t$ as $s_t = (C_t, x_t, h_t)$, which includes the current context, buffer state and channel state. At time slot $t$, the wireless user decides how many packets should be transmitted from each DU in the current context. The decision is represented by $y_t(C_t, x_t, h_t) = \{y_{f,t} \mid f \in C_t\}$, where $y_{f,t}$ represents the amount of packets transmitted from DU $f$ and $0 \leq y_{f,t} \leq x_{f,t}$. We know that the decision in time slot $t$ is a function of the current state $s_t$. In this paper, we consider the following utility[7] at each time slot $t$:

$$u(s_t, y_t) = \sum_{f \in C_t} q_f y_{f,t} - \lambda \rho\left(h_t, \sum_{f \in C_t} y_{f,t}\right) \quad (1)$$

In this utility function, the first term represents the distortion reduction obtained by transmitting the amount of data $y_t = [y_{f,t}]_{f \in C_t}$ from the DUs in the current context. The second term represents the negative value of the consumed energy by transmitting the amount $\sum_{f \in C_t} y_{f,t}$ of packets, where $\lambda > 0$ is the parameter trading-off the distortion reduction and the consumed energy. The energy consumption function $\rho(h,y)$ is assumed to be a convex function of $y$ given the channel condition $h$. One example is $\rho(h,y) = \sigma^2(e^{2y}-1)/h$, which is derived from the information-theoretic rate-power function [20]. Then, the wireless user aims to maximize the following long-term expected discounted utility:

$$\max_{y_t(s_t),\forall t} \mathbf{E}\left\{\sum_{t=0}^{\infty} \alpha^t u(s_t, y_t)\right\} \quad (2)$$

where $\alpha \in [0,1)$ is the discount factor. Note that when $\alpha \to 1$, the optimal solution to the optimization in (2) is equivalent to the optimal solution to the problem maximizing the average utility. For independently decodable DUs, given the decision $y_t(s_t)$ in time slot $t$, the buffer state transition is

$$x_{f,t+1} = \begin{cases} x_{f,t} - y_{f,t} & \text{if } f \in C_t \cap C_{t+1} \\ l_f & \text{if } f \in C_{t+1} \setminus C_t \end{cases}, \quad (3)$$

where the notation "$C_t \cap C_{t+1}$" represents the set of DUs persist from time slot $t$ to time slot $t+1$ (i.e. do not expired at the end of time slots $t$) and $C_{t+1} \setminus C_t$ represents the set of DUs that arrive in time slot $t+1$.

---

[7] In this paper, we consider that the multimedia quality is defined as the total distortion reduction of the received media packets.

From the above discussion, we know that the channel state transition and buffer state transition are Markovian. We further notice that the buffer state transition also depends on $y_t$, which is the decision made by the wireless user. Hence, the transition of the state $s_t = (C_t, x_t, h_t)$ is Markovian and the problem in Eq. (2) can be formulated as an MDP [14]. In the subsequent sections, we will discuss how the packet scheduling problem can be solved using an MDP formulation.

*A. MDP formulation and post-decision state-based dynamic programming*

In the problem in Eq. (2), the decision in each time slot $t$ is to determine the amount of data, $y_{f,t}$ to be transmitted for each DU $f \in C_t$. From [14], we know that the optimal decision can be found by solving the following Bellman's equations:

$$V(C_t, x_t, h_t) = \max_{0 \leq y_t \leq x_t} \left\{ \sum_{f \in C_t} q_f y_{f,t} - \lambda \rho\left(h_t, \sum_{f \in C_t} y_{f,t}\right) + \alpha \mathbf{E}_{h_{t+1}|h_t, l_{t+1}} V(C_{t+1}, (x_t - y_t) \oplus l_{t+1}, h_{t+1}) \right\}, \quad (4)$$

where $V(C_t, x_t, h_t)$ is the state-value function representing the optimal long-term utility starting from the state $(C_t, x_t, h_t)$ and $l_{t+1} = \{l_{f'}\}_{f' \in C_{t+1} \setminus C_t}$. The operator $z_t \oplus l_{t+1}$ represents $\{z_{f,t}\}_{f \in C_t \cap C_{t+1}} \cup \{l_{f'}\}_{f' \in C_{t+1} \setminus C_t}$ where $\{z_{f,t}\}_{f \in C_t \cap C_{t+1}}$ represents the remaining data (i.e. data which was not transmitted at time slot $t$) in DU $f \in C_t \cap C_{t+1}$ after the data transmission at time slot $t$ and $\{l_{f'}\}_{f'\in C_{t+1}\setminus C_t}$ represents the newly arriving data in DU $f' \in C_{t+1} \setminus C_t$ at time slot $t+1$. It is easy to see that the buffer state fulfils the following condition: $x_{t+1} = (x_t - y_t) \oplus l_{t+1}$. The expectation in Eq. (4) is taken over all the possible new channel states $h_{t+1}$ with the probability of $p(h_{t+1}|h_t)$ and over the possible data arrival for the DUs in the set $C_{t+1} \setminus C_t$ with the probability of $\prod_{f' \in C_{t+1} \setminus C_t} PMF_{f'}(l_{f'})$. Note that the context transition in Eq. (4) is deterministic.

From Eq. (4), it is worth to note that the expectation (over the data arrival and channel state transition) is embedded into the term to be maximized. However, in a real system, the distribution of the data arrival for each DU and channel state transition is often unavailable a priori, which makes it computationally impossible to directly optimize the long-term utility shown in Eq. (2). Similar to [21], we introduce an intermediate state which represents the state after transmitting the data (making the decision), but before the new data arrives and new channel state is realized. This intermediate state is referred to as the post-decision state $\tilde{s}_t$. In order to differentiate the "post-decision" state $\tilde{s}_t$ from the state $s_t$, we refer to the state $s_t$ as the "normal" state. The post-decision state at time slot $t$ is also illustrated in Figure 2. From this figure, we know that the post-decision state is a deterministic function of the normal state $s_t$ and the decision $y_t$, which is given by $\tilde{s}_t = (C_t, z_t, h_t)$, where $z_t = x_t - y_t$.

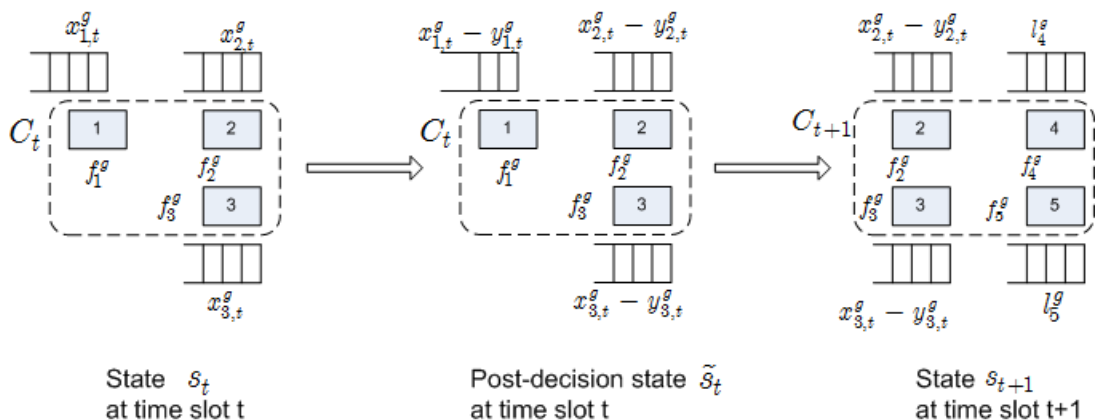

Figure 2. Post-decision state illustration

Similarly, we introduce the post-decision state-value function $U(C_t, z_t, h_t)$ to represent the optimal long-term utility starting from the post-decision state $\tilde{s}_t = (C_t, z_t, h_t)$. Then, we can rewrite the Bellman's equations in Eq. (4) as:

$$U(C_t, z_t, h_t) = \mathbf{E}_{h_{t+1}|h_t, l_{t+1}} V(C_{t+1}, z_t \oplus l_{t+1}, h_{t+1}) \text{ and} \quad (5)$$

$$V(C_t, \bm{x}_t, h_t) = \max_{0 \leq \bm{y}_t \leq \bm{x}_t} \left\{ \sum_{f \in C_t} q_f y_{f,t} - \lambda \rho \left( h_t, \sum_{f \in C_t} y_{f,t} \right) + \alpha U(C_t, \bm{x}_t - \bm{y}_t, h_t) \right\} \quad (6)$$

The first equation shows that the post-decision state-value function $U$ is obtained from the normal state-value function $V$ by taking the expectation over the possible data arrivals and possible channel transitions. The second equation shows that the normal state-value function is then obtained from the post-decision state-value function $U$ by performing the maximization over the possible decision, which is referred to as the *foresighted decision* since the optimal decision is found by maximizing the long-term utility. However, when performing the foresighted decision illustrated in Eq. (6), we face the following challenges:

- At each time slot, there are multiple DUs that are available for transmission. Determining the amount of data to transmit for each DU is a multivariable optimization which is often too complicated to solve at each time slot. However, it is fortunate that the DUs can be prioritized based on their heterogeneous data features. This prioritization will allow us to separate the multi-DU foresighted decision in Eq. (6) into multiple single-DU foresighted decisions (which is single-variable optimization). The separation will be presented in Sections III.C.

- In video transmission systems, we do not know the statistical knowledge of the underlying dynamics (e.g. channel state transition, the amount of packets for newly arriving DUs). However, after introducing the post-decision state $\tilde{s}_t = (C_t, \bm{z}_t, h_t)$, we can separate the media transmission system into two phases: the foresighted decision phase, which is governed by Eq. (6) and the dynamic realization phase, which is governed by Eq. (5). We further notice that, given the post-decision state-value function $U$, the foresighted decision phase is independent of the dynamic realization phase. This motivates us to directly learn the post-decision state-value function when the underlying dynamics are unknown. In Section III.D, we will present how the post-decision state-value functions can be learned over time for the separated foresighted decisions.

*B. Transmission priority of DUs*

In this section, we aim to define the transmission priorities between DUs. At each time slot $t$, the optimal amount of data to be transmitted from DU $f \in C_t$ is denoted by $y_{f,t}^*$.

**Definition (Transmission priority):** At any time slot $t$, if $f, f' \in C_t$ and $\left(x_{f,t} - y_{f,t}^*\right) y_{f',t}^* = 0$ for any $\boldsymbol{x}_t \geq 0$ and any channel state $h_t$, then DU $f$ has a higher transmission priority than DU $f'$, which is denoted by $f \triangleleft f'$.

The above definition on the priority indicates that, when DU $f$ has a higher transmission priority than DU $f'$, then the data from DU $f$ will be transmitted before the data from DU $f'$ is transmitted. Given the optimal post-decision state value function $U(C_t, \boldsymbol{x}_t, h_t)$, we can prioritize the DUs as follows.

**Lemma** 1 (**Prioritization using optimal post-decision state value functions**): For any two DUs $f, f' \in C_t$, if

$$U\left(C_t, \boldsymbol{x} + e_f, h_t\right) - U\left(C_t, \boldsymbol{x} + e_{f'}, h_t\right) < \left(q_f - q_{f'}\right)/\alpha, \forall \boldsymbol{x}, \qquad (7)$$

where $e_f$ is a vector which has the same dimension as $\boldsymbol{x}$ and the element corresponding to DU $f \in C_t$ is 1 and the elements corresponding to all other DUs are 0, then $f \triangleleft f'$.

*Proof*: see Appendix A.

This lemma shows that, if the optimal post-decision state value function $U(C_t, \boldsymbol{x}_t, h_t)$ satisfies the inequality in Eq. (7) at the current context $C_t$ and channel state $h_t$, then the optimal decision is to transmit the data from DU $f$ before the data from DU $f'$, which means that DU $f$ has a higher transmission priority than DU $f'$.

However, as shown in Lemma 1, in order to determine the priorities of the DUs, we have to compute the optimal post-decision state value function first, which may not be possible in practical video transmission systems since we cannot obtain the post-decision state value function without first solving the Bellman's equations in Eqs. (5) and (6). However, we are able to derive the priorities between the DUs based on the heterogeneous attributes of DUs without computing the optimal post-decision state value functions.

**Lemma** 2 (**Prioritization using the heterogeneous attributes of independent DUs**): For DUs $f, f' \in C_t$, if $q_f \geq q_{f'}$ and $d_f \geq d_{f'}$ (equalities do not hold at the same time), then $f \triangleleft f'$.

*Proof*: see Appendix B.

The priority $f \triangleleft f'$ indicates that $\left(x_{f,t} - y_{f,t}^*\right) y_{f',t}^* = 0$ at any time slot $t$ when $f, f' \in C_t$. This further implies that: (i) The buffer state $x_{f',t}$ of DU $f'$ does not affect the decision on the amount of data, $y_{f,t}^*$, to be transmitted from DU $f$ at any time slot $t$; (ii) When starting to transmit the data from DU $f'$, all the data from DU $f$ must be transmitted, i.e. the post-decision traffic state for DU $f$ is zero. In the next section, we will utilize the priorities between DUs and present the separation in the multi-DU foresighted decision given in Eq. (6) and develop a low-complexity scheduling algorithm.

*C. Priority graph-assisted scheduling*

Given the transmission priority between DUs derived based on the DUs' attributes as shown in Lemma 2, we are able to construct a DAG to represent the priorities of the DUs at each time slot, which is referred to as the priority graph and denoted by $PG_t = \langle C_t, E_t \rangle$, where $C_t$ is the set of DUs available for transmission and $E_t$ is the set of edges representing the priorities between two DUs. In this priority graph, if $f \triangleleft f'$, then there is an edge in $E_t$ pointed from DU $f'$ to DU $f$. Two examples of priority graphs are shown in Figure 3. Note that the priority graph is different from the dependency graph [7], which is built only based on the source coding dependencies between DUs.

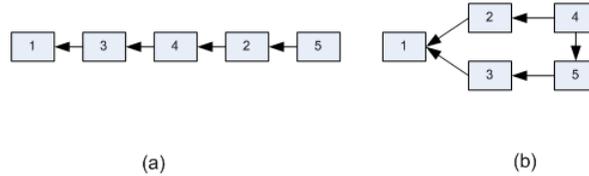

(a)      (b)

Figure 3. Examples of priority graphs with five DUs: (a) all DUs are prioritized (chain); (b) only part of the DUs are prioritized (e.g. the priorities of DUs 2 and 3 are unknown).

In the following, we will try to separate the multi-DU foresighted decision in Eq. (6) into multiple one-DU foresighted decisions. We first consider that the DUs available for transmission at each time slot can be fully prioritized (i.e. the corresponding priority graph is a chain as shown in Figure 3 (a)). The following theorem shows that we can decompose the multi-DU foresighted decision into multiple single-DU foresighted decisions at each time slot.

**Theorem** 1 **(Separation principle for multi-DU foresighted decision with priority graph of chain)**: When the DUs in each context are prioritized as a chain, then the optimal decision for each DU at each time slot can be computed as

$$y_{f,t}^* = \arg\max_{0 \leq y_{f,t} \leq x_{f,t}} \left\{ q_f y_{f,t} - \lambda \rho \left( h_t, \sum_{f' \triangleleft f} x_{f',t} + y_{f,t} \right) + \alpha U_f \left( C_t, x_{f,t} - y_{f,t}, h_t \right) \right\}, \quad (8)$$

where the post-decision state value function $U_f(C_t, x, h_t)$ satisfies the following Bellman's equations:

if $f \in C_t \cap C_{t+1}$, then

$$U_f(C_t, x_{f,t}, h_t) = \mathop{\mathbf{E}}_{h_{t+1}|h_t} \mathop{\mathbf{E}}_{l_{f'}: f' \triangleleft f, f' \in C_{t+1}/C} \max_{0 \leq y_{f,t+1} \leq x} \left\{ q_f y_{f,t+1} - \lambda \rho \left( h_{t+1}, \sum_{\forall f' \triangleleft f, f' \in C_{t+1}/C_t} l_{f'} + y_{f,t} \right) + \alpha U_f \left( C_{t+1}, x_{f,t} - y_{f,t}, h_{t+1} \right) \right\} \quad (9)$$

and if $f \in C_t / C_{t+1}$, then $U_f(C_t, x, h_t) = 0$.

*Proof*: See Appendix C.

*Remark 1*: Theorem 1 indicates that, we can find the optimal decision $y_{f,t}^*$ for each DU $f \in C_t$ by solving the foresighted decision given in Eq. (8) from the highest priority DU to the lowest priority DU. In this foresighted decision, the post-decision state-value function $U_f(C_t, x_{f,t} - y_{f,t}, h_t)$ only depends on the buffer state $x_{f,t} - y_{f,t}$ of DU $f$ and is independent of the buffer states of the other DUs in the current context $C_t$. This is because when transmitting the data from DU $f$, the data from DU $f'(f' \triangleleft f)$ has been transmitted (i.e. the buffer state is zero), and the data from DU $f'(f \triangleleft f')$ cannot be transmitted (i.e. the buffer state will not affect the foresighted decision in Eq. (8)). We note that, when making the foresighted decision for DU $f$, the transmission cost is $\rho \left( h_t, \sum_{f' \triangleleft f} x_{f',t} + y_{f,t} \right) - \rho \left( h_t, \sum_{f' \triangleleft f} x_{f',t} \right)$, which is the marginal transmission cost of transmitting the data from DU $f$. However, the term $\rho \left( h_t, \sum_{f' \triangleleft f} x_{f',t} \right)$ is independent of the decision variable $y_{f,t}$ and hence, it is not shown in the foresighted decision in Eq. (8).

*Remark 2*: The post-decision state value function for DU $f$ is computed as in Eq. (9). If DU $f$ is expired at time slot $t+1$, then the post-decision state value function is zero, otherwise it is computed by solving the Bellman's equations for DU $f$. In the Bellman's equations, we note that the post-decision state-value function for DU $f$ is not affected by DUs $f'(f \triangleleft f')$. In fact, it only depends on the buffer states of the DUs that arrive at time slot $t+1$ and have a higher priority than DU $f$. We can also note that the update of the post-decision state-value function for DU $f'$ is not affected by DU $f$.

We now consider a general scenario where the priorities of DUs at each time slot are represented by a general priority graph instead of a chain. The priority graph for the DUs at time slot $t$ is given by $PG_t = \langle C_t, E_t \rangle$. Similar to Theorem 1, for any two DUs $f, f' \in C_t$, if $f \triangleleft f'$, then we should transmit the data from DU $f$ first and the buffer state of DU $f'$ does not affect the foresighted decision for DU $f$. However, the buffer state of DU $f$ will affect the transmission cost of DU $f'$ in the foresighted decision, but will not affect the update of the post-decision state value function. If $f$ and $f'$ are not prioritized, we have to decide which DU should be transmitted first and how much data should be transmitted from this selected DU. The following theorem answers this question.

**Theorem** 2 (**Separation principle for multi-DU foresighted decision with general priority graph**): Given the priority graph $PG_t = \langle C_t, E_t \rangle$ at time slot $t$, the optimal decisions for the DUs are performed as in Algorithm 1. After determining the optimal decisions, the post-decision state value function of DU $f$ is updated as follows.

If $f \in C_t \cap C_{t+1}$, then

$$U_f(C_t, x_{f,t}, h_t) = \mathop{\mathbf{E}}_{h_{t+1}|h_t} \mathop{\mathbf{E}}_{l_{f'}: f' \tilde{\triangleleft} f, f' \in C_{t+1}/C} \max_{0 \leq y_{f,t+1} \leq x_{f,t}} \left\{ q_f y_{f,t+1} - \lambda \rho \left( h_{t+1}, \sum_{f' \tilde{\triangleleft} f, f' \in C_{t+1}/C_t} l_{f'} + y_{f,t+1} \right) + \alpha U_f(C_{t+1}, x_{f,t} - y_{f,t+1}, h_{t+1}) \right\} \quad (10)$$

else, $U_f(C_t, x, h_t) = 0$, where $f' \tilde{\triangleleft} f$ [8] means that the transmission order of DU $f'$ is earlier than DU $f$, which is determined in Algorithm 1.

Algorithm 1: Optimal packet scheduling induced by the priority graph for independent DUs

---
**Input**: $PG_t$, $U_f(C_t, x_{f,t} - y_{f,t}, h_t)$
**Initialize**: $PG_t^0 = PG_t$
**For** $k = 1, \cdots, |C_t|$:

$$f^k = \arg \max_{f \in root(PG_t^k)} \max_{0 \leq y_{f,t} \leq x_{f,t}} \left\{ q_f y_{f,t} - \lambda \rho \left( h_t, \sum_{j=1}^{k-1} y_{f^j,t}^* + y_{f,t} \right) + \alpha U_f(C_t, x_{f,t} - y_{f,t}, h_t) \right\} \quad (11)$$

$$y_{f^k,t}^* = \arg \max_{0 \leq y_{f^k,t} \leq x_{f^k,t}} \left\{ q_{f^k} y_{f^k,t} - \lambda \rho \left( h_t, \sum_{j=1}^{k-1} y_{f^j,t}^* + y_{f^k,t} \right) + \alpha U_{f^k}(C_t, x_{f^k,t} - y_{f^k,t}, h_t) \right\} \quad (12)$$

where $root(PG_t^k)$ is the operator extracting the roots of the priority graph $PG_t^k$ and $PG_t^k = PG_t^{k-1}/\{f^{k-1}\}$.
**Return** $(f^1, \cdots, f^{|C_t|})$ and $(y_{f^1,t}^*, \cdots, y_{f^{|C_t|},t}^*)$.
---

*Proof*: The proof can be derived similarly to that of Theorem 1.

The optimal packet scheduling illustrated in Algorithm 1 can be easily explained as follows. Starting from the priority graph $PG_t$, we compare the DUs that are the roots in the priority graph $PG_t^k$ and select the DU with the highest long-term utility to transmit as shown in Eq. (11). The optimal scheduling for the selected DU is found by solving the corresponding foresighted decision as shown in Eq. (12). Finding the optimal packet scheduling in state $s_t = (C_t, \boldsymbol{x}_t, h_t)$ as illustrated in Algorithm 1 can also be interpreted by using a priority tree, which is uniquely constructed from the priority graph $PG_t$ corresponding to the context $C_t$. Two examples of priority trees, which correspond to the priority graphs in Figure 3, are given in Figure 4. The root of the priority tree is the priority graph $PG_t$ and each node is a priority graph. The child nodes of each node in the priority tree are obtained by removing one of the root packets in the priority graph at this node. Then, finding the optimal packet scheduling is equivalent to "travelling" the priority tree induced by the priority graph $PG_t$.

The update of the post-decision state-value function $U_f(C_t, x, h_t)$ for DU $f$ is performed independently of the other DUs in the current context $C_t$ as shown in Eq. (10), which is the same as the update of the post-decision state value function for the fully prioritized DUs presented in Theorem 1. However, unlike in Theorem 1, when updating the post-decision state-value function $U_f(C_t, x, h_t)$, we cannot directly prioritize all the DUs arriving at time slot $t+1$, i.e. the DUs in the set of $C_{t+1} \setminus C_t$ because we may not be able to compare the transmission priority between the arriving DUs with the DUs persisting from time slot $t$, i.e. in the set of $C_{t+1} \cap C_t$. Hence, we resort to the priority graph built for the DUs in the set of $C_{t+1} \setminus C_t \cup \{f\}$ and Algorithm 1 in order to determine the transmission orders of the DUs in this set. When the transmission order of the DUs $C_{t+1} \setminus C_t \cup \{f\}$ is determined, we can update the post-decision state-value function $U_f(C_t, x, h_t)$ for DU $f$ which only depends on the DUs that are transmitted before it (i.e. DUs $f'$ such that $f' \tilde{\triangleleft} f$) at the same time slot. It is easy to show that Algorithm 1 preserves the transmission priority (i.e. DUs with higher priorities are always transmitted before DUs with lower priorities).

---

[8] $f' \triangleleft f$ implies that $f' \tilde{\triangleleft} f$ but not vice versa.

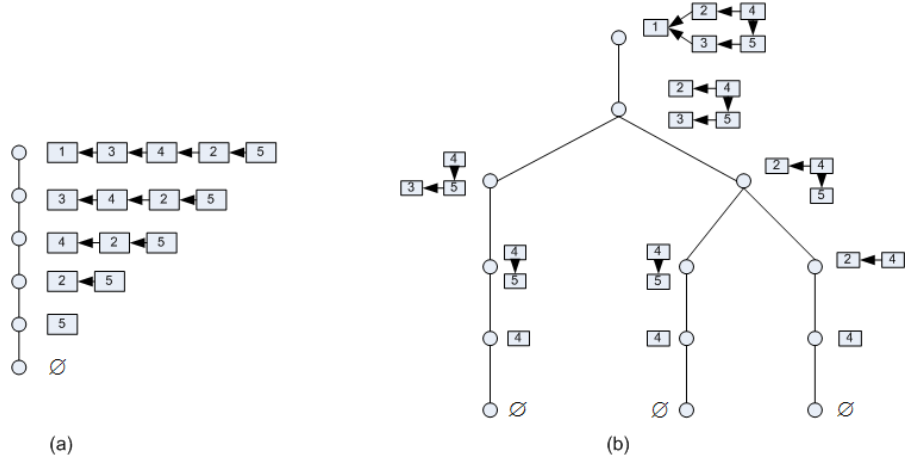

(a)                          (b)

Figure 4. Priority trees induced from (a) the graph in which all packets are prioritized (chain) (b) the graph in which some packets are prioritized and others are not.

### D. Online learning

In multimedia transmission systems, we do not know the statistical knowledge about the packet arrivals for each DU and the underlying channel state transition. Without this statistical knowledge, we cannot directly update the post-decision state value function for each DU as shown in Eq. (10). In the following, we present how we can update the post-decision state-value functions for the DUs without having the statistical knowledge about the underlying dynamics.

From Eq. (10), we know that the expectation over the dynamics is performed outside of the maximization and that the dynamics (including the packet arrivals and channel state transition) are independent of the buffer states. Then, we are able to update the post-decision state-value function using time-average as presented in [22]. That is, we can update the post-decision state-value function for all the possible buffer states for each DU at each time slot. The online learning algorithm for each DU $f \in C_t$ is presented as follows. At time slot $t$, we first perform Algorithm 1 to determine the optimal decision for all the DUs in the context of $C_t$ and the transmission order $(f^1, \cdots, f^{|C_t|})$. Then the post-decision state-value function for DU $f \in C_t \cap C_{t-1}$ is updated for all the possible buffer state $x \in [0, l_f^{\max}]$ as follows:

$$U_{f,t}(C_{t-1}, x, h_{t-1}) =$$
$$(1-\beta_t) \max_{0 \le y_{f,t+1} \le x} \left\{ q_f y_{f,t} - \lambda \rho \left( h_t, \sum_{f' \tilde{\lhd} f, f' \in C_t / C_{t-1}} l_{f'} + y_{f,t} \right) + \alpha U_{f,t-1}(C_t, x - y_{f,t}, h_t) \right\} \quad (13)$$
$$+ \beta_t U_{f,t-1}(C_{t-1}, x, h_{t-1})$$

where $\beta_t$ is a diminishing step size, e.g. $\beta_t = 1/t$. As in [22], we can further show that the post-decision state-value function updated as in Eq. (13) is concave in the buffer state $x$ given the context and channel state. Then, we are able to adopt a piece-wise linear function to approximate the post-decision state-value function and perform the online learning algorithm with adaptive approximation presented in [22], which can significantly reduce the complexity of the online learning algorithm. We omit the details here and refer the interested readers to [22].

## IV. PACKET SCHEDULING FOR INTERDEPENDENT DUs

In this section, we aim to develop a packet scheduling solution for interdependent DUs. Different from the independent DUs case, due to inter-dependencies, the scheduling decision on each DU $f$ will be affected by the amount of data transmitted from the DUs on which DU $f$ depends on. As discussed in Section II, the dependency between DUs is expressed by a DAG which is different from the priority graph defined in Section III.B. In order to capture the impact of the dependency between the DUs, we introduce a dependency factor $p_f(v) \in [0,1]$ for each DU $f$ to represent the impact on the distortion reduction of those DUs that depend on DU $f$, which is a function of the amount of data remaining when DU $f$ is expired. One example of dependency factor is $p_f(z) = \exp(-\beta_f z)$ as given in [23][24]. Then, the utility at each time slot is given by

$$u(s_t, \boldsymbol{y}_t) = \sum_{f \in C_t} \prod_{f' \prec f} p_{f'}\left(z_{f',d_{f'}}\right) q_f y_{f,t} - \lambda \rho\left(h_t, \sum_{f \in C_t} y_{f,t}\right). \tag{14}$$

where $z_{f',d_{f'}}$ is the remaining amount of data in the post-decision state of DU $f'$ at time slot $d_{f'}$ (i.e. the amount of data from DU $f'$ that are not received by the decoder). The difference between Eq. (14) and Eq. (1) is that the distortion reduction in Eq. (14) depends on not only the number of packets transmitted for each DU as shown in Eq. (1), but also the dependency factors of the DUs that the current DUs depend on. The long-term utility is the same as in Eq. (2).

For the interdependent DUs, in order to capture the Markovian property of the scheduling problem, we define the state at each time slot $t$ as $s_t = (C_t, \boldsymbol{p}_t, \boldsymbol{x}_t, h_t)$ to include the current context $C_t$, buffer states $\boldsymbol{x}_t$, channel state $h_t$ and dependency factor vector $\boldsymbol{p}_t$ from the parent DUs. The dependency factor vector $\boldsymbol{p}_t$ is given by $\boldsymbol{p}_t = \left[p_{f'}\left(z_{f',d_{f'}}\right)\right]_{f' \prec f, f \in C_t}$. That is, the dependency factor vector includes all the dependency factors of the DUs that the DUs in the set $C_t$ depend on. The post-decision state is defined as the state after the scheduling decision but before the new DU arrivals and the new channel state realization. We note that the post-decision dependency factor vector is the same as $p_t$. We directly use $p_t$ as the post-decision dependency factor vector in the post-decision state. Hence, the multi-DU foresighted decision based on the post-decision states is given as follows:

$$V(C_t, \bm{p}_t, \bm{x}_t, h_t) = \max_{0 \leq \bm{y}_t \leq \bm{x}_t} \left\{ \sum_{f \in C_t} \prod_{f' \prec f} p_{f'}\left(z_{f',d_{f'}}\right) q_f y_{f,t} - \lambda \rho\left(h_t, \sum_{f \in C_t} y_{f,t}\right) + \alpha U(C_t, \bm{p}_t, \bm{x}_t - \bm{y}_t, h_t) \right\}, \quad (15)$$

where $U(C_t, \bm{p}_t, \bm{z}_t, h_t)$ is the post-decision state value function.

Similar to the independent DUs, we aim to separate the multi-DU foresighted decision in Eq. (15) into multiple single-DU foresighted decision. We can introduce the priority between interdependent DUs to differentiate the transmission orders of the DUs in the current context $C_t$. However, due to the dependency, we cannot directly apply Lemma 2 here. Instead, we can prioritize the DUs using their heterogeneous attributes in the following lemma.

**Lemma** 3 (**Prioritization using heterogeneous attributes of interdependent DUs**): For any context $C_t$, if $f, f' \in C_t$ and $f \prec f'$, then $f \lhd f'$.

*Proof*: First, we notice that, if $f \prec f'$, then $q_f \geq q_{f'}$ and $d_f \leq d_{f'}$. Furthermore, when $f \prec f'$, from (14), we know that, the gained distortion reduction from DU $f'$ (i.e. $p_{f',t} q_{f'} y_{f',t}$) is discounted by the dependency factor $p_{f',t}$ which is impacted by the amount of remaining data at DU $f$. In other words, transmitting the data from DU $f$ will always achieve higher long-term utility than transmitting the data from DU $f'$, which means $f \lhd f'$. ∎

From Lemma 3, we note that, based on the dependencies between DUs, we can construct the priority graph $PG_t = \langle C_t, E_t \rangle$ for each context $C_t$. It is clear that the priority graph $PG_t$ is the dependency graph corresponding to the DUs in the current context $C_t$. At time slot $t$, given the priority graph $PG_t$ and the dependency factors vector $\bm{p}_{f,t} = \left[p_{f'}\left(z_{f',d_{f'}}\right)\right]_{f' \prec f}$ for each DU $f \in C_t$, we can perform the foresighted decision for each DU as in Algorithm 2.

**Algorithm 2: Optimal packet scheduling induced by the priority graph for interdependent DUs**

**Input**: $PG_t$, $\boldsymbol{p}_t$, $U_f(C_t, \boldsymbol{p}_{f,t}, x_{f,t}, h_t)$

**Initialize**: $PG_t^0 = PG_t$

**For** $k = 1, \cdots, |C_t|$:

$$f^k = \arg \max_{f \in root(PG_t^k)} \max_{0 \leq y_{f,t} \leq x_{f,t}} \left\{ \mathbf{1}^T \boldsymbol{p}_{f,t} \cdot q_f y_{f,t} - \lambda \rho \left( h_t, \sum_{j=1}^{k-1} y_{f^j,t}^* + y_{f,t} \right) + \alpha U_f(C_t, \boldsymbol{p}_{f,t}, x_{f,t} - y_{f,t}, h_t) \right\} \quad (16)$$

$$y_{f^k,t}^* = \arg \max_{0 \leq y_{f^k,t} \leq x_{f^k,t}} \left\{ \mathbf{1}^T \boldsymbol{p}_{f^k,t} \cdot q_{f^k} y_{f^k,t} - \lambda \rho \left( h_t, \sum_{j=1}^{k-1} y_{f^j,t}^* + y_{f^k,t} \right) + \alpha U_{f^k}(C_t, \boldsymbol{p}_{f^k,t}, x_{f^k,t} - y_{f^k,t}, h_t) \right\} \quad (17)$$

where $root(PG_t^k)$ is the operator extracting the roots of the priority graph $PG_t^k$, $PG_t^k = PG_t^{k-1} / \{f^{k-1}\}$ and update $\boldsymbol{p}_{f,t}, f \neq f^1, \cdots, f^k$.

**Return** $(f^1, \cdots, f^{|C_t|})$ and $(y_{f^1,t}^*, \cdots, y_{f^{|C_t|},t}^*)$.

In Algorithm 2, we separate the multi-DU foresighted decision and perform it by travelling the priority tree as illustrated in Section III.C, which preserves the priorities between DUs. Similarly, we can also update the post-decision state value function $U_f(C_t, \boldsymbol{p}_{f,t}, x_{f,t}, h_t)$ as follows:

If $f \in C_{t-1} \cap C_t$ (i.e. DU $f$ is not expired at both time slots),

$$U_{f,t}(C_{t-1}, \boldsymbol{p}_{f,t-1}, x, h_{t-1}) =$$

$$(1 - \beta_t) \max_{0 \leq y_{f,t+1} \leq x} \left\{ \mathbf{1}^T \boldsymbol{p}_{f,t} \cdot q_f y_{f,t} - \lambda \rho \left( h_t, \sum_{\substack{f^i, i < k \\ f^i \in C_t / C_{t-1}}} l_{f^i} + y_{f,t} \right) + \alpha U_{f,t-1}(C_t, \boldsymbol{p}_{f,t-1}, x - y_{f,t}, h_t) \right\}. \quad (18)$$

$$+ \beta_t U_{f,t-1}(C_{t-1}, \boldsymbol{p}_{f,t-1}, x, h_{t-1})$$

If $f \in C_{t-1} \setminus C_t$ (i.e. DU $f$ is expired at time slot $t$),

$$U_{f,t}(C_{t-1}, \boldsymbol{p}_{f,t-1}, x, h_{t-1}) =$$

$$(1 - \beta_t) \sum_{f': f \prec f'} \max_{0 \leq y_{f',t+1} \leq x} \left\{ \mathbf{1}^T \boldsymbol{p}_{f',t-1} \cdot q_{f'} y_{f',t} - \lambda \rho \left( h_t, \sum_{\substack{f^i, i < k \\ f^i \in C_t / C_{t-1}}} l_{f^i} + y_{f',t} \right) + \alpha U_{f',t-1}(C_t, \boldsymbol{p}_{f',t-1}, x - y_{f',t}, h_t) \right\}. \quad (19)$$

$$+ \beta_t U_{f,t-1}(C_{t-1}, \boldsymbol{p}_{f,t-1}, x, h_{t-1})$$

When DU $f$ is not expired at both time slots $t-1$ and $t$, we update the post-decision state value function using a time-average similar to the one in Eq. (13). However, when DU $f$ is expired at time slot $t$ (i.e. $f \in C_{t-1} \setminus C_t$), due to the dependency, the post-decision state of DU $f$ at time slot $t-1$ will affect the decision of those DUs in the context $C_t$ that depend on DU $f$. Hence, the post-decision state value function of DU $f$ is updated as in Eq. (19) to take into account the dependency impact on the descendent DUs, which is different from the case of independent DUs, where the post-decision state value function is zero.

However, since the dependency factor vector $\boldsymbol{p}_{f,t}$ often has large dimensions (because the DUs in the set $C_t$ depend on many DUs) and takes real values in the range of $[0,1]$, it is difficult to compute and store the post-decision state value function directly. Instead of computing the post-decision state-value function $U_f(C_t, \boldsymbol{p}_{f,t}, \boldsymbol{z}_t, h_t)$ directly, we approximate it by $\mathbf{1}^T \boldsymbol{p}_{f,t} \cdot U_f(C_t, \boldsymbol{z}_t, h_t)$, where $U_f(C_t, \boldsymbol{z}_t, h_t)$ is the post-decision state-value function corresponding to the case that $\boldsymbol{p}_{f,t} = \mathbf{1}$ which means that all the DUs that DU $f$ depends on are successfully received. Then $U_f(C_t, \boldsymbol{z}_t, h_t)$ can be updated using Eqs. (18) and (19) by setting $\boldsymbol{p}_{f,t} = \mathbf{1}$ and $\boldsymbol{p}_{f',t} = \mathbf{1}, \forall f \prec f'$. It is clear that the approximation allows us to represent the post-decision state-value function as presented in Section III.D, which significantly reduces the dimensionality of the post-decision state-value function. This is because the dependency factor vector is not the component of the arguments in the approximated post-decision state-value function any longer.

## V.  SIMULATION RESULTS

In this section, we perform several numerical experiments to verify the performance of the proposed framework and compare with various state-of-art solutions for multimedia communications.

*A.  Performance comparison of various packet scheduling solutions for video transmission*

In this section, we compare our proposed packet scheduling solution with several start-of art solutions which only consider either the heterogeneous media characteristics or the time-varying channel conditions. In the experiment, to compress the video data, we used a scalable video coding scheme [15], which is attractive for wireless streaming applications because it provides on-the-fly application adaptation to channel conditions, support for a variety of wireless receivers with different resource capabilities and power constraints, and easy prioritization of various coding layers and video packets. We choose for this experiment three video sequences (Foreman, Coastguard and Mobile at CIF resolutions, 30 frames/second), exhibiting different motion activities. The video sequences Foreman and Coastguard are encoded at the bit rate of 512 kbps and Mobile, due to its high-frequent texture and complicated motion, is encoded at 1024kbps. In this simulation, each GOP contains 16 frames and each encoded video frame can tolerate a delay of 266ms, corresponding to the half duration of a GOP.

The energy function for transmitting the amount of $y$ (in bits) traffic at the channel state $h$ is given by $c(h,y) = \sigma^2(2^y - 1)/|h|^2$, where $\sigma^2$ is the variance of the white Gaussian noise [20]. In this simulation, we choose $\bar{h}^2/\sigma^2 = 0.14$, where $\bar{h}$ is the average channel gain. We divide the entire channel gain range into eight regions each of which is represented by a representative state. The states are presented in Table 2. We choose $\alpha = 0.95$. The transmission system is time-slotted with the time slot length of 10ms.

Table 2. Channel states used in the simulation

| Channel gain ($h^2/\sigma^2$) regions | Representative states |
| --- | --- |
| (0, 0.0280], (0.0280, 0.0580] (0.0580, 0.0960] (0.0960, 0.1400] (0.1400, 0.1980] (0.1980, 0.2780], (0.2780, 0.4160] (0.4160, $\infty$ ] | 0.0131, 0.0418, 0.0753, 0.1157, 0.1661, 0.2343, 0.3407, 0.6200 |

We consider three comparable solutions: (i) our proposed packet scheduling solution which takes into account both the heterogeneous multimedia traffic characteristics (e.g. delay deadlines, distortion impacts and dependencies etc.) and time-varying network conditions; (ii) the packet scheduling solution [6] which only considers the distortion impact of each media packet and the observed channel conditions and is referred to as "distortion-impact" driven packet scheduling; (iii) the packet scheduling solution obtained by solving the rate-distortion optimization assuming the constant channel conditions (i.e. using average channel conditions) and linear transmission cost as in RaDiO [7], which is referred to as the rate-distortion optimized packet scheduling.

In Figure 5 (a)~(c), , we show the Peak-Signal-to-Noise Ratio (PSNR) as a function of the consumed energy curves under the different scheduling solutions for the three video sequences. From these figures, we note that our proposed cross-layer optimization solution outperforms both the "distortion-impact" driven packet scheduling and rate-distortion optimized packet scheduling by, on average, around 2dB and 5dB in "Foreman", 1.5dB and 3.5dB in "Coastguard", and 0.5dB and 2.5dB in "Mobile" in terms of PSNR. The improvement comes from the fact that our proposed solution schedules the packets based on the heterogeneous characteristics of the multimedia packets as well as the time-varying channel conditions. We also notice that the "distortion-impact"-driven solution obtains higher received video quality than the rate-distortion optimized packet scheduling. It shows that the time-varying channel conditions and the characteristics (dependencies, distortion impacts and delay constants) of media packets play a very important role in improving the media quality.

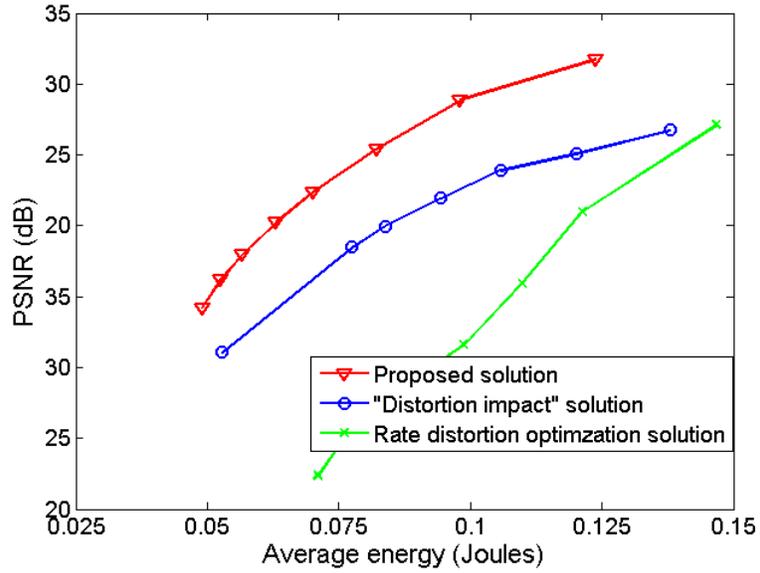

(a) "Foreman"

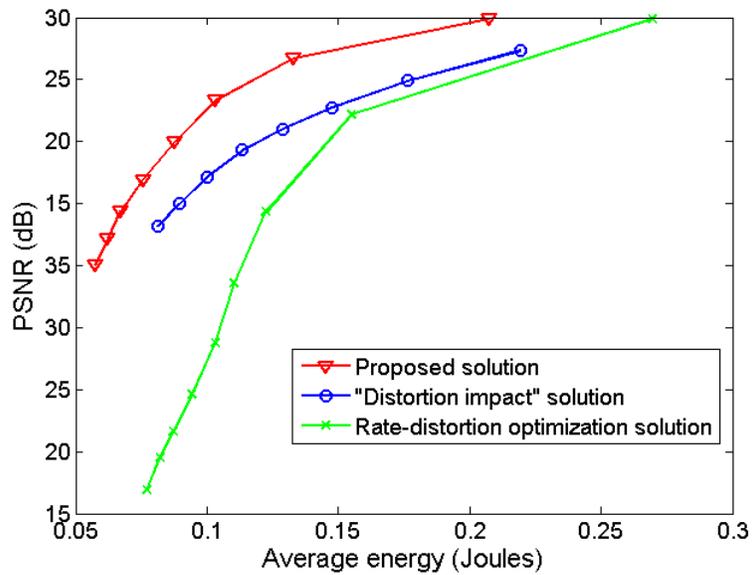

(b) "Coastguard"

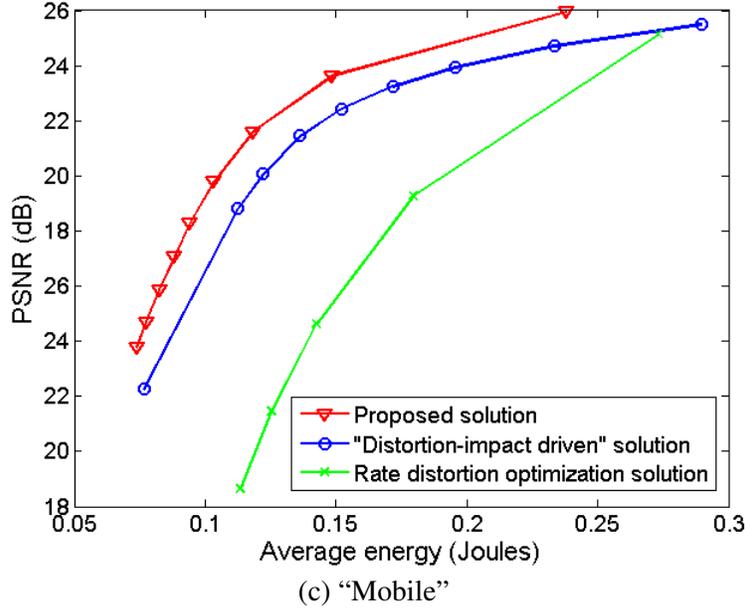
(c) "Mobile"
Figure 5. PSNR-energy curve of "Foreman", "Coastguard" and "Mobile" sequences for different transmission solutions

*B. Performance of packet scheduling optimization with various delay constraints and GOP structures*

In this section, we further compare the performance of the packet scheduling optimization solutions for streaming the Coastguard video sequence with various delay constraints and GOP structures. The wireless channel settings are the same as in Section V.A. We compare our solution with different combinations of delay deadlines and GOP structures. The PSNR versus consumed energy curves are given in Figure 6 and Figure 7. From Figure 6, we notice that, when the video sequence is encoded with the GOP of 16 frames, by increasing the delay from 266 ms to 533 ms, the packet scheduling optimization can improve, on average, 1 dB in terms of PSNR. From Figure 7, we further notice that, when the video sequence is encoded with the GOP of 8 frames, by increasing the delay from 133ms to 266ms, the packet scheduling optimization can improve, on average, 1.5 dB in terms of PSNR. The improvement comes from the fact that, by increasing the delay, each media packet has more transmission opportunities and will be scheduled for transmission when a better channel condition is encountered. By increasing the number of frames in one GOP, the video sequence can be encoded more efficiently and there are fewer packets to be transmitted, which accordingly improves the video quality.

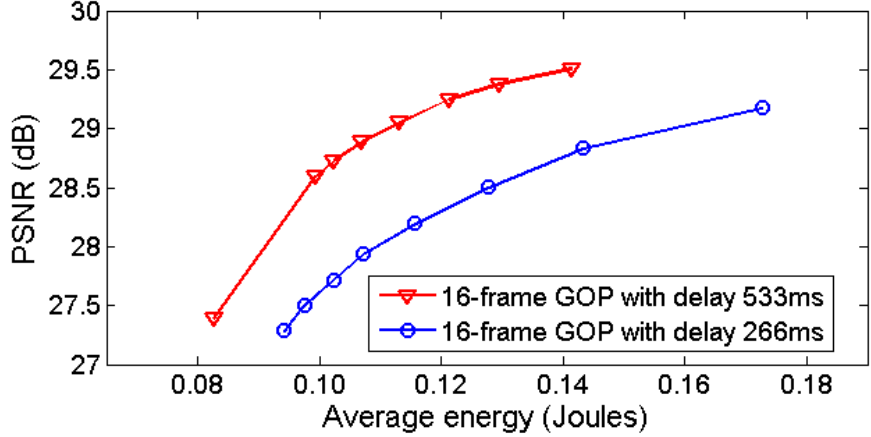

Figure 6.  Video quality of "Coastguard" sequence with various delay deadlines when 16-frame GOP is used

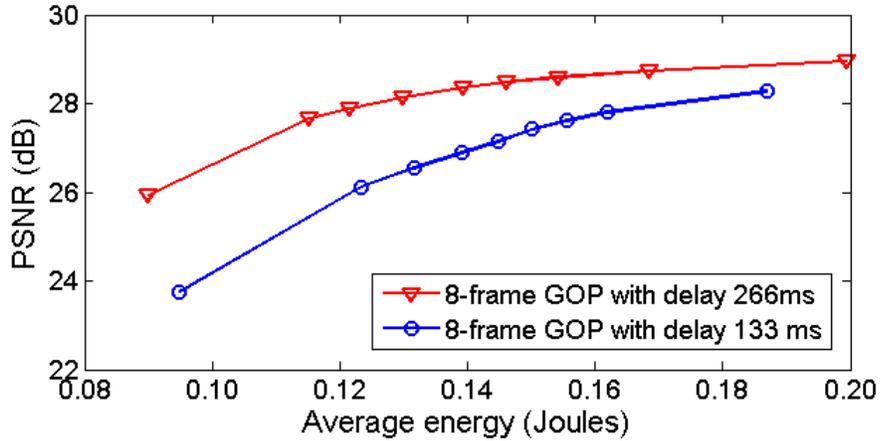

Figure 7.  Video quality of "Coastguard" sequence with various delay deadlines when 8-frame GOP is used

## VI.  CONCLUSIONS

In this paper, we formulate the problem of packet scheduling optimization for delay-sensitive packetized media applications as a Markov decision process. Based on the heterogeneous characteristics of the media packets, we express the transmission priorities between DUs as a DAG. Using the DAG expression, we are able to separate the multi-data unit foresighted decision at each time slot into multiple single-data unit foresighted decisions, which can subsequently be performed from the high priority DU to the low priority DU. The post-decision state-value function associated with each DU is updated individually using the online learning algorithms. The simulation results show that the proposed foresighted optimization solution significantly outperforms the start-of-art solutions which (partially) ignore the media characteristics and time-varying network conditions. This proposed systematic scheduling framework is general and can be easily applied to many other multimedia-related problems. For example, using the proposed context to represent the heterogeneous video data encoded by different video codecs, e.g. H.264 [16], SVC [15][17], etc., the proposed packet scheduling can also be applied to energy-efficient video encoding/decoding systems with dynamic voltage scaling [27], by separating the multi-DU scheduling decision into multiple single-DU scheduling decisions. When the transmission acknowledgement is delayed, as it is the case for multi-hop wired and wireless networks, we can easily extend the proposed scheduling framework using a partially-observed MDP formulation and then apply the proposed separation for the foresighted decision. We further notice that the packet scheduling algorithm can also be deployed in the middle nodes of a multi-hop networks (e.g. mesh or sensor networks) to relay the multimedia data from other nodes.

## APPENDIX

### A. Proof of Lemma 1:

*Proof*: The optimal decision at time slot $t$ is denoted by $\boldsymbol{y}_t^* = \left(y_{f,t}^*, y_{f',t}^*, \boldsymbol{y}_{-f-f',t}^*\right)$ where $\boldsymbol{y}_{-f-f',t}^*$ represents the optimal decision for the DUs other than DUs $f, f'$. We assume that $\left(x_{f,t} - y_{f,t}^*\right) y_{f',t}^* \neq 0$, which means that $x_{f,t} - y_{f,t}^* > 0$ and $y_{f',t}^* > 0$. We consider another decision $\boldsymbol{y}_t = \left(y_{f,t}^* + 1, y_{f',t}^* - 1, \boldsymbol{y}_{-f-f',t}^*\right)$ which is feasible since $x_{f,t} - y_{f,t}^* > 0$ and $y_{f',t}^* > 0$. We compare the long-term utility associated with the decision $\boldsymbol{y}_t$ to the one associated with the optimal decision $\boldsymbol{y}_t^*$ (i.e. $V(C_t, \boldsymbol{x}_t, h_t)$) as follows.

$$-q_f \left(x_{f,t} - y_{f,t}^* - 1\right) - q_{f'} \left(x_{f',t} - y_{f',t}^* + 1\right) - \sum_{f'' \in C_t \setminus C_{t+1}} q_{f''} \left(x_{f'',t} - y_{f'',t}^*\right) - \lambda \rho \left(h_t, \sum_{f'' \in C_t} y_{f'',t}\right)$$
$$+ \alpha U\left(C_t, \boldsymbol{x} - \boldsymbol{y}_t^* - \boldsymbol{e}_f + \boldsymbol{e}_{f'}, h_t\right)$$
$$= q_f - q_{f'} + \alpha U\left(C_t, \boldsymbol{x} - \boldsymbol{y}_t^* - \boldsymbol{e}_f + \boldsymbol{e}_{f'}, h_t\right) - \alpha U\left(C_t, \boldsymbol{x} - \boldsymbol{y}_t^*, h_t\right) + V(C_t, \boldsymbol{x}_t, h_t)$$
$$> V(C_t, \boldsymbol{x}_t, h_t)$$

The inequality is due to the fact that

$$q_f - q_{f'} + \alpha U\left(C_t, \boldsymbol{x} - \boldsymbol{y}_t^* - \boldsymbol{e}_f + \boldsymbol{e}_{f'}, h_t\right) - \alpha U\left(C_t, \boldsymbol{x} - \boldsymbol{y}_t^*, h_t\right)$$
$$= q_f - q_{f'} + \alpha U\left(C_t, \boldsymbol{x}' - \boldsymbol{y}_t^* + \boldsymbol{e}_{f'}, h_t\right) - \alpha U\left(C_t, \boldsymbol{x}' - \boldsymbol{y}_t^* + \boldsymbol{e}_f, h_t\right) > 0$$

where $\boldsymbol{x}' = \boldsymbol{x}' - \boldsymbol{e}_f$ and the inequality is from the condition given in Eq. (7).

Hence, $\left(x_{f,t} - y^*_{f,t}\right)y^*_{f',t} \neq 0$ implies that $y^*_t$ is not the optimal decision which contradicts the assumption. ∎

*B. Proof of Lemma 2:*

*Proof*: To prove this, we only need to show that

$$U\left(C_t, \boldsymbol{x} + e_f, h_t\right) - U\left(C_t, \boldsymbol{x} + e_{f'}, h_t\right) < \left(q_f - q_{f'}\right)/\alpha, \forall \boldsymbol{x}, t \leq d_f.$$

We prove this using backward induction.

At time slot $t = d_f$, on the one hand, we have $U(C_t, \boldsymbol{x} + e_f, h_t) = U(C_t, \boldsymbol{x}, h_t)$ because DU $f$ will expire and be deleted at the next slot and have no contribution to the post-decision state value function. On the other hand, we can prove that $0 \leq U\left(C_t, \boldsymbol{x} + e_{f'}, h_t\right) - U(C_t, \boldsymbol{x}, h_t) \leq q_{f'}$ because the best utility we can obtain in the future time slots by transmitting one packet in DU $f'$ is $q_{f'}$. Then

$$-q_{f'} \leq U\left(C_t, \boldsymbol{x} + e_f, h_t\right) - U\left(C_t, \boldsymbol{x} + e_{f'}, h_t\right)$$
$$= U(C_t, \boldsymbol{x}, h_t) - U\left(C_t, \boldsymbol{x} + e_{f'}, h_t\right) \leq 0 < \left(q_f - q_{f'}\right)/\alpha.$$

Now, we assume that, at time slot $t \leq d_f$, $U\left(C_t, \boldsymbol{x} + e_f, h_t\right) - U\left(C_t, \boldsymbol{x} + e_{f'}, h_t\right) < \left(q_f - q_{f'}\right)/\alpha, \forall \boldsymbol{x}$. We try to prove that $U\left(C_{t-1}, \boldsymbol{x} + e_f, h_{t-1}\right) - U\left(C_{t-1}, \boldsymbol{x} + e_{f'}, h_{t-1}\right) < \left(q_f - q_{f'}\right)/\alpha, \forall \boldsymbol{x}$ at time slot $t$. From the Bellman's equation in Eq. (5), we know that $U\left(C_{t-1}, \boldsymbol{x}_{t-1} + e_f, h_{t-1}\right) = \mathbf{E}V\left(C_t, \boldsymbol{x}_t + e_f, h_t\right)$ where $\boldsymbol{x}_{t+1}$ is derived from $\boldsymbol{x}_t$ by deleting the expired DUs and adding the new arriving DUs. Hence, it is equivalent to prove that $V\left(C_t, \boldsymbol{x} + e_f, h_t\right) - V\left(C_t, \boldsymbol{x} + e_{f'}, h_t\right) < \left(q_f - q_{f'}\right)/\alpha, \forall \boldsymbol{x}$. We denote the optimal scheduling decision in computing $V(C_t, \boldsymbol{x}, h_t)$ by $\boldsymbol{y}^*_t(\boldsymbol{x})$. Then the optimal decision in computing $V\left(C_t, \boldsymbol{x} + e_f, h_t\right)$ can be in three cases: (1) $\boldsymbol{y}^*_t\left(\boldsymbol{x} + e_f\right) = \boldsymbol{y}^*_t(\boldsymbol{x})$, i.e. the additional packet in DU $f$ is not transmitted; (2) $\boldsymbol{y}^*_t\left(\boldsymbol{x} + e_f\right) = \boldsymbol{y}^*_t(\boldsymbol{x}) + e_f - e_{f''}$, i.e. transmitting the additional packet in DU $f$ instead of the packet in DU $f''$; (3). $\boldsymbol{y}^*_t\left(\boldsymbol{x} + e_f\right) = \boldsymbol{y}^*_t(\boldsymbol{x}) + e_f$, i..e transmitting the additional packet in DU $f$. Similarly, the optimal decision in computing $V\left(C_t, \boldsymbol{x} + e_{f'}, h_t\right)$ has also three cases. However, we have the following relationship: if $\boldsymbol{y}^*_t\left(\boldsymbol{x} + e_f\right)$ is case $i = 1, 2, 3$, then $\boldsymbol{y}^*_t\left(\boldsymbol{x} + e_{f'}\right)$ should be case $i' = 1, \cdots, i$. For all the cases, we can prove that $V\left(C_{t+1}, \boldsymbol{x}_{t+1} + e_f, h_{t+1}\right) - V\left(C_{t+1}, \boldsymbol{x}_{t+1} + e_{f'}, h_{t+1}\right) \leq q_f - q_{f'}/\alpha$. For example, we consider that $\boldsymbol{y}^*_t\left(\boldsymbol{x} + e_f\right)$ is case 3 and $\boldsymbol{y}^*_t\left(\boldsymbol{x} + e_{f'}\right)$ is case 1. Then

$$V\left(C_t, \boldsymbol{x} + e_f, h_t\right) = \sum_{f'' \in C_t} q_{f''} y^*_{f'',t} + q_f - \lambda \rho \left( \sum_{f'' \in C_t} q_{f''} y^*_{f'',t} + 1, h_t \right) + \alpha U\left(C_t, \boldsymbol{x} - \boldsymbol{y}^*_t(\boldsymbol{x}), h_t\right) \quad \text{and}$$

$$V\left(C_t, \boldsymbol{x} + e_{f'}, h_t\right) = \sum_{f'' \in C_t} q_{f''} y^*_{f'',t} - \lambda \rho \left( \sum_{f'' \in C_t} q_{f''} y^*_{f'',t}, h_t \right) + \alpha U\left(C_t, \boldsymbol{x} - \boldsymbol{y}^*_t(\boldsymbol{x}) + e_{f'}, h_t\right). \text{ It is clear that}$$

$$\begin{aligned}
V\left(C_t, \boldsymbol{x} + e_{f'}, h_t\right) &= \sum_{f'' \in C_t} q_{f''} y^*_{f'',t} - \lambda \rho \left( \sum_{f'' \in C_t} q_{f''} y^*_{f'',t}, h_t \right) + \alpha U\left(C_t, \boldsymbol{x} - \boldsymbol{y}^*_t(\boldsymbol{x}) + e_{f'}, h_t\right) \\
&\geq \sum_{f'' \in C_t} q_{f''} y^*_{f'',t} + q_{f'} - \lambda \rho \left( \sum_{f'' \in C_t} q_{f''} y^*_{f'',t} + 1, h_t \right) + \alpha U\left(C_t, \boldsymbol{x} - \boldsymbol{y}^*_t(\boldsymbol{x}), h_t\right)
\end{aligned},$$

since $\boldsymbol{y}^*_t(\boldsymbol{x} + e_{f'})$ is in case 3. Hence, we can have $V\left(C_t, \boldsymbol{x} + e_f, h_t\right) - V\left(C_t, \boldsymbol{x} + e_{f'}, h_t\right) \leq q_f - q_{f'}$.

We can similarly prove the other scenarios. Finally, we prove that $U\left(C_{t-1}, \boldsymbol{x} + e_f, h_{t-1}\right) - U\left(C_{t-1}, \boldsymbol{x} + e_{f'}, h_{t-1}\right) < \left(q_f - q_{f'}\right)/\alpha$ ∎.

## C. Proof of Theorem 1:

Proof: Since the DUs at each time slot are prioritized as a chain, from Lemma 1, we know that, the optimal decision for each DU is found, starting from the highest priority DU, by solving the following optimization.

$$y^*_{f,t} = \arg \max_{0 \leq y_{f,t} \leq x_{f,t}}$$

$$\left\{ q_f y_{f,t} - \lambda \rho \left( h_t, \sum_{f' \triangleleft f, f' \in C_t} x_{f',t} + y_{f,t} \right) + \alpha U\left(C_t, \{0_{f',t}\}_{f' \triangleleft f} \cup \{x_{f,t} - y_{f,t}\} \cup \{x_{f',t}\}_{f \triangleleft f'}, h_t\right) \right\}.$$

where $0_{f',t}$ represents that DU $f'$ is empty. As we know, when performing the foresighted decision for DU $f$, all the data from DU $f'$ $(f' \triangleleft f)$ has been transmitted and no data from DU $f'$ $(f \triangleleft f')$ is transmitted. It is true for any time slot. Hence, we are able to split the post-decision state-value function in the above foresighted optimization into two parts:

$$U\left(C_t, \{0_{f',t}\}_{f' \triangleleft f} \cup \{x_{f,t} - y_{f,t}\} \cup \{x_{f',t}\}_{f \triangleleft f'}, h_t\right) = U_f\left((C_t, \{x_{f,t} - y_{f,t}\}, h_t)\right) + U_{\{f \triangleleft f'\}}\left(C_t, \{x_{f',t}\}_{f \triangleleft f'}, h_t\right)$$

where $U_f\left((C_t, \{x_{f,t} - y_{f,t}\}, h_t)\right)$ represents the long-term utility obtained for DU $f$ and $U_{\{f \triangleleft f'\}}\left(C_t, \{x_{f',t}\}_{f \triangleleft f'}, h_t\right)$ represents the long-term utility obtained for all the DUs $f'(f \triangleleft f')$. The reason that we can split it is that, the lower priority DU $f'(f \triangleleft f')$ will not affect the foresighted decision for DU $f$ and the data from the lower priority DU can be transmitted only if the DUs from the higher priority DUs have been transmitted (i.e. empty) at the current time slot. Hence, $U_{\{f \triangleleft f'\}}\left(C_t, \{x_{f',t}\}_{f \triangleleft f'}, h_t\right)$ only depends on the amount of data from DU $f$ ($f \triangleleft f'$) observed before the foresighted decision (i.e. $x_{f,t}$) and is independent of the decision at current time slot. Hence, the foresighted decision for DU $f$ can be rewritten as

$$y_{f,t}^* = \arg\max_{0 \leq y_{f,t} \leq x_{f,t}} \left\{ q_f y_{f,t} - \lambda \rho \left( h_t, \sum_{f' \triangleleft f, f' \in C_t} x_{f',t} + y_{f,t} \right) + \alpha U_f\left(C_t, x_{f,t} - y_{f,t}, h_t\right) \right\}$$

which is the form given in Eq. (8) and $U_f\left(C_t, \tilde{x}_{f,t}, h_t\right)$ is the post-decision state value function associated with DU $f$.

The update of the post-decision state-value function can be shown using backward induction, as shown in [22]. We refer the interested reader to the proof in [22] for more details.